\documentclass[twocolumn,journal]{IEEEtran}

\usepackage{graphicx}
\usepackage{amsmath}
\usepackage{amssymb}
\usepackage{comment}

\usepackage{amsthm}
\usepackage{cite}
\usepackage{float}
\usepackage{color}
\usepackage{balance}

    \newtheorem{remark}{Remark}%
\begin{document}
\title{\LARGE Security of Rechargeable Energy-Harvesting\\  Transmitters in Wireless Networks}
\author{Ahmed El Shafie$^\dagger$, Dusit Niyato$^\star$, Naofal Al-Dhahir$^\dagger$
\begin{tabular}{c}
\small $^\dagger$Electrical Engineering Dept., University of Texas at Dallas, USA. \\
\small $^\star$School of Computer Engineering, Nanyang Technological University (NTU), Singapore.
\thanks{This paper was made possible by NPRP grant number 6-149-2-058 from the Qatar National Research Fund (a member of Qatar Foundation). The statements made herein are solely the responsibility of the authors.}
\end{tabular}
\thanks{This paper is published in IEEE Wireless Communications Letters.}
\vspace{-2\baselineskip} 
}
\date{}
\maketitle
\begin{abstract}
In this letter, we investigate the security of a single-antenna rechargeable source node in the presence of a multi-antenna rechargeable cooperative jammer and a potential single-antenna eavesdropper. The batteries at the legitimate transmitting nodes (i.e. the source node and the jamming node) are assumed to be limited in capacity and are modeled as queueing systems. We investigate the impact of the energy arrival rates at the batteries on the achievable secrecy rates. In our energy-constrained network, we propose an efficient scheme to enhance the system's security by optimizing the transmission times of the source node. The jammer uses a subset of its antennas (and transmit radio-frequency chains) to create a beamformer which maximizes the system's secrecy rate while completely canceling the artificial noise at the legitimate destination. Our numerical results demonstrate the significant average secrecy rate gain of our proposed scheme.
\end{abstract}
\vspace{-0.6cm}
\section{Introduction}
\vspace{-0.1cm}
In battery-based energy-constrained communication systems, network lifetime maximization
is very crucial \cite{6951347,van2004prolonging}. Energy-harvesting
schemes were integrated into communication systems as a powerful solution
to the problem of limited network lifetime since terminals can harvest
energy from ambient energy sources (solar, wind, etc.) \cite{alippi2008adaptive}.

Security is critical for wireless channels due to the broadcast nature of the medium.
In \cite{5278549}, the authors assumed a source node (Alice) that wishes to communicate with her destination node (Bob) in the presence of a multi-antenna friendly jammer (Jimmy) and an eavesdropping node (Eve). The jammer was assumed to transmit artificial noise (AN) to maximize the secrecy rate. Moreover, the eavesdropper's channel state information (CSI) was assumed perfectly known at the legitimate nodes. The optimal beamforming (BF) vector and power allocation at the jammer were designed to enhance the system's secrecy rate. In \cite{mukherjee2012deploying}, the authors proposed the deployment of an energy-harvesting jammer in a multiple-input multiple-output wiretap channel. The authors assumed that the jamming signal vector is not orthogonal to the Alice-Bob channel vector.

Motivated by \cite{5278549} and \cite{mukherjee2012deploying}, we consider the impact of transmitting nodes' batteries on the security of the wireless network in \cite{5278549} when both Alice and Jimmy are equipped with limited-capacity rechargeable batteries. The batteries are charged by the energy harvested from nature.

The contributions of this letter are summarized as follows. \begin{itemize}
    \item  We investigate the network in \cite{5278549} when both Alice and Jimmy are equipped with limited-capacity rechargeable batteries. We investigate the impact of the energy arrival rates at the batteries on the system's secrecy rate.
        \item Instead of using all antennas at Jimmy for jamming Eve as in \cite{5278549}, we propose to use a subset of Jimmy's antennas for jamming. In addition, we optimize the data transmission time to further improve the secrecy rate.
        \item We show that when one of the two batteries is saturated with energy, the other battery is modeled as a Geo/Geo/1 queueing system. We also investigate the well-known Geo/D/1 with \emph{unity service rate} queueing model for nature energy-harvesting systems \cite{krikidis,ourletter}, which generally achieves a lower-bound on the actual system performance. This lower-bound enables us to relate the average arrival rates at the batteries with the achievable secrecy rate.
        \end{itemize}
\vspace{-0.4cm}
\section{System Model and Assumptions}
\vspace{-0.1cm}
We consider the following network model. A source node (Alice) communicates with her destination node (Bob) in the presence of a friendly jamming node (Jimmy) and an eavesdropping node (Eve). Similar to the model in \cite{5278549}, Alice, Bob and Eve have single antennas while Jimmy has $\mathcal{N}$ antennas labeled as $1,2,\dots,\mathcal{N}$. We denote Alice, Bob, Eve and Jimmy by ${\rm A}$, ${\rm B}$, ${\rm E}$, and ${\rm J}$, respectively. Time is partitioned into equal-size time slots whose duration is $T$ time units and the channel has a bandwidth of $W$ Hz. We assume flat-fading channels. The channel coefficient between Node $n$ and Node $m$, denoted by $h_{n,m}$, remains constant during a time slot, but it changes identically and independently (i.i.d.) from one time slot to another. For Jimmy, we use an integer number to indicate the antenna index. The thermal noise at a receiving node is modeled as an additive white Gaussian noise (AWGN) with zero mean and variance $\kappa$ Watts/Hz.

 We assume that Alice and Jimmy are energy-harvesting nodes with energy batteries modeled as queueing systems as in, e.g., \cite{krikidis,ourletter} and the references therein. The energy arrivals at Node $k\in \{\rm A,J\}$ are i.i.d. Bernoulli random variables with average $\lambda_k$ energy packets/slot \cite{krikidis,ourletter}.\footnote{Although we assume i.i.d. energy arrivals at each node as in \cite{krikidis,ourletter}, the case of correlated arrivals at the nodes can be considered in all parts of this letter. However, we need this assumption to analyze the energy queues Markov chains in Section \ref{queueinganalysis2} and Appendix \ref{markov}.} The Bernoulli arrival model is simple, but it still
can capture the random and sporadic nature of packet arrival at the batteries. The battery at Node $k\in \{\rm A,J\}$ is denoted by $B_k$ and has a maximum capacity of $\mathcal{B}_k^{\max}$.

Assuming the energy arrival model in \cite{krikidis,ourletter}, each energy packet arrives with certain amount of energy and is transmitted with the same amount of energy. We assume that an energy packet at Alice contains $e_{\rm A}$ energy units and at Jimmy contains $e_{\rm J}$ energy units. When Node $k$ transmits, its average transmit power is $e_{k}/T_{k}$, where $T_k$ is the transmission time. The AN signals used in jamming are modeled as zero-mean circularly-symmetric complex Gaussian random variables~\cite{5278549}.
\vspace{-0.3cm}
\section{Proposed Jamming Scheme}
\vspace{-0.1cm}
The secrecy outage happens when the transmission rate exceeds the secrecy rate. Letting $\mathcal{C}^{\mathcal{L}}_{n,m}$ denote the channel capacity of the $n-m$ link when the event $\mathcal{L}$ is true, the secrecy rate of the Alice-Bob link is given by
 \begin{equation} \small
 \label{okpp}
C^{\mathcal{L}}_{{\rm s},{\rm A}}\!=\left[\mathcal{C}^\mathcal{L}_{{\rm A},{\rm B}}-\mathcal{C}^\mathcal{L}_{{\rm A},{\rm E}} \right]^+ \le \mathcal{C}^\mathcal{L}_{{\rm A},{\rm B}}=\mathcal{C}_{{\rm A},{\rm B}}
\end{equation}
where $[\cdot]^+$ denotes the maximum between the enclosed value between brackets and \emph{zero} and $\mathcal{L}\in\{\{B_{\rm J}>0\}, \{B_{\rm J}=0\}\}$ represents the state of Jimmy's battery. If $\mathcal{L}=\{B_{\rm J}>0\}$ ($\mathcal{L}=\{B_{\rm J}=0\}$), Jimmy's battery has (no) energy and hence he can(not) help in jamming Eve. The last equality in (\ref{okpp}) follows from the fact that the Alice-Bob link rate does not change with Jimmy's activity.\footnote{The battery state (i.e. empty or nonempty) at Alice and Jimmy can be announced to all nodes using a known pilot.}

Our proposed jamming scheme is summarized as follows
\begin{itemize}
\item In each time slot, if Alice and Jimmy batteries have energy, Alice transmits her data with rate equal to the secrecy rate $C^{B_{\rm J}>0}_{{\rm s},{\rm A}}$. We assume that during Alice's transmission, Jimmy creates a beamformer to maximize the secrecy rate of Alice while completely canceling the AN interference at Bob. The weights used at Jimmy are chosen to null the interference at Bob while maximizing the interference at Eve's receiver.
\item If Alice's battery has energy and Jimmy's battery has no energy (hence he cannot transmit the AN signal), Alice transmits her data with secrecy rate $C^{B_{\rm J}=0}_{{\rm s},{\rm A}}$.
\item If Alice's battery has no energy, she cannot transmit data and hence she and Jimmy remain idle during the current time slot.
    \end{itemize}
     A similar BF-jamming scheme was proposed in \cite{5278549}. However, our approach is distinct in the following aspects: 1) Instead of using all of Jimmy's antennas for jamming Eve, which requires $\mathcal{N}$ radio-frequency (RF) chains, we assume that only a set of $\mathcal{K}$ RF chains is available at Jimmy (or he only activates any $\mathcal{K} \le \mathcal{N}$ of them during the transmissions).\footnote{For simplicity, we assume that antennas labeled from $1$ to $\mathcal{K}$ are used at Jimmy for jamming Eve.} This reduces the power consumption and hardware design complexity since the scheme reduces the number of RF chains and antennas to $\mathcal{K}$ and also reduces signal processing complexity since we need to estimate fewer channels to apply BF-jamming. 2) We optimize Alice's transmission times to enhance the achievable secrecy rate due to the increase of the transmit and jamming powers. In addition, we derive closed-form expressions for the optimal weight vector at Jimmy using a geometric method of
orthogonal projection. Moreover, we obtain expressions for the system's secrecy rate and its average. 3) We analyze the energy arrival randomness at Alice and Jimmy and show their impact on the average secrecy rate.

We start by investigating the case when both Alice and Jimmy are active. Let $\mathcal{J} \in\{1,2,\dots,\mathcal{K}\}$ with cardinality $\mathcal{K}\le \mathcal{N}$ denote the set of Jimmy's antennas that are used to jam Eve. Jimmy designs a cooperative beamformer using his antennas in $\mathcal{J}$ to maximize the secrecy rate of Alice. Full CSI is assumed at all nodes including Eve's CSI as in \cite{5278549}. This assumption is valid when Eve is an active node in the network, i.e., another node that communicates with Bob.

Let $\Gamma_{\rm A}=\frac{e_{\rm A}}{T}$ and $\Gamma_{\rm J}=\frac{e_{\rm J}}{T}$. For given channel realizations, the rates of the Alice-Bob and Alice-Eve links are
 \begin{equation} \small
 \begin{split}
 \label{okkkooo}
\mathcal{C}_{{\rm A},{\rm B}}\!&=\!\alpha_{\rm A}\log_2\left(1\!+\!\frac{\frac{\Gamma_{\rm A}}{\alpha_{\rm A}}\theta_{{\rm A},{\rm B}}}{\kappa W}\right)\!\\ \mathcal{C}^{B_{\rm J}>0}_{{\rm A},{\rm E}}\!&=\alpha_{\rm A}\log_2\left(\!1\!+\!\frac{\frac{\Gamma_{\rm A}}{\alpha_{\rm A}} \theta_{{\rm A},{\rm E}}}{\kappa W\!+\!\frac{\Gamma_{\rm J}}{\alpha_{\rm A}} |\sum_{j\in \mathcal{J}} g_j^* h_{j,{\rm E}}|^2 }\right)
\end{split}
\end{equation}
where $\alpha_{\rm A}=T_{\rm A}/T \in[0,1]$. The secrecy rate is $C^{B_{\rm J}>0}_{\rm s,A}=[\mathcal{C}_{{\rm A},{\rm B}}-\mathcal{C}^{B_{\rm J}>0}_{{\rm A},{\rm E}}]^+$ and a positive secrecy rate is achieved when $\frac{\theta_{{\rm A},{\rm B}}}{\kappa W}> \frac{\theta_{{\rm A},{\rm E}}}{\kappa W+\frac{\Gamma_{\rm J}}{\alpha_{\rm A}} |\sum_{j\in \mathcal{J}} g_j^* h_{j,{\rm E}}|^2 }$. The superscript $^*$ denotes the complex-conjugate transpose, $|\cdot|$ denotes the absolute value, $\theta_{j,k}=|h_{j,k}|^2$ denotes channel gain (i.e. squared magnitude of the channel coefficient $h_{j,k}$) between Node $j\in\{{\rm A},1,2,3,\dots,\mathcal{N}\}$ and Node $k\in\{\rm E,B,1,2,\dots,\mathcal{N}\}$, and $\mathbf{g}=[g_1,g_2,\dots,g_{\mathcal{K}}]^\top$, where the superscript $^\top$ denotes vector transpose, is the BF weight vector whose dimension is $\mathcal{K} \times 1$ with $g_j$ as the weight used at Antenna~$j\in \mathcal{J}$.

 From (\ref{okkkooo}), the signal-to-interference-plus-noise ratio (SINR) at Bob increases with $\alpha_{\rm A}$ while the numerator and denominator of the SINR at Eve increases with $\alpha_{\rm A}$. Hence, the appropriate selection of $\alpha_{\rm A}$ can enhance the secrecy rate.

We aim at maximizing the secrecy rate in a given time slot over the weight vector used at Jimmy and the transmission time. That is,
 \begin{equation} \small
 \label{ccccc}\small
\!\! \underset{\substack{{\mathbf{g}}\\{\alpha_{\rm A}\!\in [0,1]}}}{\max:}  \alpha_{\rm A}\!\left[\!\log_2\!\left(\!\!1\!+\!\frac{\frac{\Gamma_{\rm A}}{\alpha_{\rm A}}\theta_{{\rm A},{\rm B}}}{\kappa W}\!\right) \!-\!\log_2\!\left(\!1\!+\!\frac{\frac{\Gamma_{\rm A}}{\alpha_{\rm A}} \theta_{{\rm A},{\rm E}}}{\kappa W\!\!+\!\frac{\Gamma_{\rm J}}{\alpha_{\rm A}} |\!\sum_{j\!\in \mathcal{J}} g_j^* h_{j,{\rm E}}|^2 }\!\right)\!\!\right]\!\!.
\end{equation}
For a given (fixed) $\alpha_{\rm A}$, we notice that the optimization problem becomes independent of $\alpha_{\rm A}$. This implies that the optimal weight vector is independent of $\alpha_{\rm A}$. Hence, we can solve two separate optimization problems. More specifically, for a fixed $\alpha_{\rm A}$, maximizing $C^{B_{\rm J>0}}_{{\rm s},{\rm A}}$ over the weight vector $\mathbf{g}$ is equivalent to minimizing $\log_2\left(1+\frac{\frac{\Gamma_{\rm A}}{\alpha_{\rm A}} \theta_{{\rm A},{\rm E}}}{\kappa W+\frac{\Gamma_{\rm J}}{\alpha_{\rm A}} |\sum_{j\in \mathcal{J}} g_j^* h_{j,{\rm E}}|^2 }\right)$. Since the logarithmic function is a monotonically increasing function, the problem reduces to the maximization of the following objective function
 \begin{equation} \small
 \label{cccccx}
\!\underset{\mathbf{g}}{\max:} \ C^{B_{\rm J>0}}_{{\rm s},{\rm A}}\!
\rightarrow \underset{\mathbf{g}}{\max:} \ { |\sum_{j\in \mathcal{J}} g_j^* h_{j,{\rm E}}|^2 }.
\end{equation}
The simplified objective function, $ { |\sum_{j\in \mathcal{J}} g_j^* h_{j,{\rm E}}|^2 }$, is completely independent of $\alpha_{\rm A}$.

Let $\mathbf{h}_{\rm E}\!=\![h_{1,{\rm E}},h_{2,{\rm E}},\dots,h_{\mathcal{K},{\rm E}}]^\top\in \mathbb{C}^{\mathcal{K}\times 1}$ denote the coefficients vector from the Jimmy's antennas to Eve's antenna, where $\mathbb{C}^{\mathcal{K}\times 1}$ denotes the set of all ${\mathcal{K}}$-dimensional complex vectors, and $\mathcal{K}$ represents the number of used antennas in jamming Eve. Moreover, $\mathbf{h}_{\rm B}=[h_{1,{\rm B}},h_{2,{\rm B}},\dots,h_{\mathcal{K},{\rm B}}]
^\top \in \mathbb{C}^{\mathcal{K}\times 1}$ denotes the coefficients vector from Jimmy's antennas to Bob's antenna. The optimal weight vector $\mathbf{g}$ that maximizes $|\mathbf{g}^* \mathbf{h}_{\rm E}|^2=|\sum_{j\in \mathcal{J}} g_j^* h_{j,{\rm E}}|^2$ subject to (${\rm s.t.}$) the normalization constraint $\|g\|^2=1$, where $\|\cdot\|$ represents the $\ell_2$-norm, and the total elimination of the interference at Bob, i.e., $|\mathbf{g}^* \mathbf{h}_{\rm B}|=0$, can be achieved by solving the following optimization problem
\begin{eqnarray}
\label{opt2x}
\small \begin{split} \small
  \! \underset{\mathbf{g}}{\max:}  & \,\,\,\,\  |\mathbf{g}^* \mathbf{h}_{\rm E}|^2, \ {\rm s.t.}    \,\,\,\,\     |{\mathbf{g}}^* \mathbf{h}_{\rm B}| \!=\! 0, \ \|\mathbf{g}\|^2\!=\!1.
     \end{split}
\end{eqnarray}
 Since both the objective function and the constraints are independent of $\alpha_{\rm A}$, the optimal weight vector is independent of $\alpha_{\rm A}$ as mentioned earlier. To solve this problem, we first note that the optimal weight vector must null the interference at Bob. This implies that the optimal weight vector is orthogonal to $\mathbf{h}_{\rm B}$ and belongs to a subspace orthogonal to the channel vector $\mathbf{h}_{\rm B}$. Let $\mathbb{V}$ denote the orthogonal complementary
subspace of the subspace spanned by $\mathbf{h}_{\rm B}$. Then, we choose the weight vector that belongs to $\mathbb{V}$ and at the same time maximizes the term  $|\mathbf{g}^* \mathbf{h}_{\rm E}|^2$.  According to the closest point theorem \cite{cpt}, the optimal weight vector is the orthogonal
projection of $\mathbf{h}_{\rm E}$ onto the subspace $\mathbb{V}$. Since $\mathbf{g}$ has a unit norm, we must divide the projection vector by its magnitude. Thus,
\begin{equation}
\small
\mathbf{g}^\star= \frac{\Psi \mathbf{h}_{\rm E}}{\|\Psi \mathbf{h}_{\rm E}\|}
\end{equation}
 where $\Psi$ is the projection matrix which is given by $\Psi\!=\!\mathbf{I}_\mathcal{K}\!-\!\frac{\mathbf{h}_{\rm B}{\mathbf{h}_{\rm B}}^*}{{\|\mathbf{h}_{\rm B}}\|^2}$, and $\mathbf{I}_\mathcal{K}$ denotes the identity matrix whose size is $\mathcal{K}\times \mathcal{K}$. Then, we substitute with $\mathbf{g}=\mathbf{g}^\star$ into the objective function of \eqref{ccccc} and optimize \eqref{ccccc} over $\alpha_{\rm A}$.
\begin{remark}
\label{remark1}
If Eve's CSI is unknown at the legitimate nodes, Jimmy designs the AN vector to lie in a subspace orthogonal to the subspace spanned by the channel vector between Jimmy and Bob. In this case, the optimal beamformer is a precoding matrix, denoted by ${\mathbf{G}}$, and is given by the solution of $ \mathbf{h}^\top_{\rm B} {\mathbf{G}}=0$. The columns of ${\mathbf{G}} \in \mathbb{C}^{\mathcal{K} \times (\mathcal{K}-1)}$ are combined using an AN vector of zero-mean circularly-symmetric complex Gaussian random variables. Since the AN precoding matrix has $\mathcal{K}-1$ columns, the AN vector size is $(\mathcal{K}-1)\times 1$.
\end{remark}
Finally, we investigate the case when Alice's battery has energy and Jimmy's battery has no energy. When Jimmy's battery is empty, the secrecy rate is given by
\begin{equation} \small
\!\! C^{B_{\rm J}=0}_{{\rm s},{\rm A}}\!=\!\alpha_{\rm A}\left[\!\log_2\left(1\!+\!\frac{\frac{\Gamma_{\rm A}}{\alpha_{\rm A}}\theta_{{\rm A},{\rm B}}}{\kappa W}\right)\!-\log_2\left(\!1\!+\!\frac{\frac{\Gamma_{\rm A}}{\alpha_{\rm A}} \theta_{{\rm A},{\rm E}}}{\kappa W\! }\right)\!\right]^+ \!\!\!\!\le \! C^{B_{\rm J}>0}_{{\rm s},{\rm A}}
\end{equation}
with positive secrecy rate when $\theta_{{\rm A},{\rm B}}\!>\!  \theta_{{\rm A},{\rm E}}$.
\vspace{-0.18cm}
\section{Batteries Queueing Analyses}
\vspace{-0.01cm}
\label{queueinganalysis}
Let $\overline{C^{B_{\rm J}=0}_{{\rm A},{\rm B}}}=\mathbb{E}\{C^{B_{\rm J}=0}_{{\rm A},{\rm B}}\}$ and $\overline{C^{B_{\rm J}>0}_{{\rm A},{\rm B}}}=\mathbb{E}\{C^{B_{\rm J}>0}_{{\rm A},{\rm B}}\}$ denote the average secrecy rate of Alice transmission when Jimmy has no energy and has energy to help, respectively, where $\mathbb{E}\{\cdot\}$ denotes the statistical expectation. When Jimmy's battery is empty and Alice's battery is nonempty, Alice transmits with secrecy rate $C^{B_{\rm J}=0}_{{\rm s},{\rm A}}$. When Jimmy's battery is nonempty and Alice's battery is nonempty, Alice transmits with secrecy rate $C^{B_{\rm J}>0}_{{\rm s},{\rm A}}$. Hence, the average number of securely decoded bits/sec/Hz at Bob is given by
 \begin{equation} \small
\begin{split} \small
\label{mua}
\!\! \mu_{\rm A}\!=\!  \left(\overline{C^{B_{\rm J}>0}_{{\rm A},{\rm B}}} \Pr\{B_{\rm A}\!>\!0,B_{\rm J}\!>\!0\} \!+\!\overline{C^{B_{\rm J}=0}_{{\rm A},{\rm B}}} \Pr\{B_{\rm A}\!>\!0,B_{\rm J}\!=\!0\}\right).
\end{split}
\end{equation}

        When $\frac{\theta_{{\rm A},{\rm B}}}{\kappa W}\le  \frac{ \theta_{{\rm A},{\rm E}}}{\kappa W+\frac{\Gamma_{\rm J}}{\alpha_{\rm A}} |\sum_{j\in \mathcal{J}} g_j^* h_{j,{\rm E}}|^2 }$ or Alice's battery is empty, there is no energy leaving Jimmy's battery. Hence, the average service rate of $B_{\rm J}$ is
         \begin{equation} \small
         \label{la1}
\mu_{B_{\rm J}}\!=\! \Pr\{B_{\rm A}>0\}\Pr\left\{\frac{\theta_{{\rm A},{\rm B}}}{\kappa W}> \frac{ \theta_{{\rm A},{\rm E}}}{\kappa W+\frac{\Gamma_{\rm J}}{\alpha_{\rm A}} |\sum_{j\in \mathcal{J}} g_j^* h_{j,{\rm E}}|^2 }\right\}.
\end{equation}

       An energy packet is depleted from Alice's battery when Jimmy's battery is nonempty and the channel is secure or when Jimmy's battery is empty and the channel is secure. Hence, the average service rate of Alice's battery is given by
         \begin{equation}
         \begin{split}
         \label{la2}
\mu_{B_{\rm A}}&\!=\! \Pr\{B_{\rm J}>0\} \beta+ \Pr\{B_{\rm J}=0\} \Pr\left\{\theta_{{\rm A},{\rm B}}>  \theta_{{\rm A},{\rm E}}\right\}
\end{split}
\end{equation}
where $\beta=\Pr\left\{\frac{\theta_{{\rm A},{\rm B}}}{\kappa W}>\frac{\theta_{{\rm A},{\rm E}}}{\kappa W+\frac{\Gamma_{\rm J}}{\alpha_{\rm A}} |\sum_{j\in \mathcal{J}} g_j^* h_{j,{\rm E}}|^2 }\right\}$ and $\frac{\theta_{{\rm A},{\rm B}}}{\kappa W}> \frac{ \theta_{{\rm A},{\rm E}}}{\kappa W+\frac{\Gamma_{\rm J}}{\alpha_{\rm A}} |\sum_{j\in \mathcal{J}} g_j^* h_{j,{\rm E}}|^2 }$ and $\theta_{{\rm A},{\rm B}}>\theta_{{\rm A},{\rm E}}$ are the conditions to achieve a positive secrecy rate when Jimmy's battery is nonempty and empty, respectively.

From \eqref{la1} and \eqref{la2}, the service processes of Alice and Jimmy batteries are coupled and the battery states are correlated. Hence, it is not possible to obtain closed-form expressions for the marginal and joint probabilities in $\mu_{\rm A}$, $\mu_{B_{\rm J}}$, and $\mu_{B_{\rm A}}$. Nevertheless, in the following subsections, we investigate two important special cases to gain some insights.
\vspace{-0.3cm}
\subsection{The case of large batteries capacities and $\lambda_{\rm A}\!=\!1$ or $\lambda_{\rm J}\!=\!1$}
\label{queueinganalysis2}
\subsubsection{The case of $\lambda_{\rm A}=1$} When $\lambda_{\rm A}=1$, Alice always has energy to transmit data. In other words, she has a reliable energy supply. Hence, $\Pr\{B_{\rm A}=0\}=0$ and $\Pr\{B_{\rm A}>0\}=1$. The average service rates of the energy queues are thus given by $\mu_{B_{\rm J}}= \beta$ and
         \begin{equation}
         \begin{split}
\!\mu_{B_{\rm A}}&\!=\! \Pr\{B_{\rm J}\!>\!0\} \beta \!+\! \Pr\{B_{\rm J}=0\} \Pr\left\{{\theta_{{\rm A},{\rm B}}}\!>\!{ \theta_{{\rm A},{\rm E}}}\right\}.
\end{split}
\end{equation}
Moreover, the average secrecy rate is given by
\begin{equation} \small
\begin{split} \small
\label{popp}
\mu_{\rm A}\!=\!  \overline{C^{B_{\rm J}>0}_{{\rm A},{\rm B}}} \Pr\{B_{\rm J}>0\} +\overline{C^{B_{\rm J}=0}_{{\rm A},{\rm B}}} \Pr\{B_{\rm J}=0\}.
\end{split}
\end{equation}
Since the average service rate of $B_{\rm J}$ does not depend on the state of $B_{\rm A}$, and the arrival process is stationary with average $\lambda_{\rm J}$, $B_{\rm J}$ becomes a Geo/Geo/1/$\mathcal{B}_{\rm J}^{\max}$. We analyze its Markov chain in Appendix \ref{markov}. When $\mathcal{B}_{\rm J}^{\max}$ is very large, the probability that Jimmy's battery is nonempty is given by
   \begin{equation} \small
   \label{jimy}
\Pr\{ B_{\rm J} >0\}= \min\left\{\frac{\lambda_{\rm J}}{\beta},1\right\}.
\end{equation}
Substituting with (\ref{jimy}) into (\ref{popp}), the average secrecy rate of the system is given by
\begin{equation} \small
\begin{split} \small
\mu_{\rm A}\!=\!  \overline{C^{B_{\rm J}>0}_{{\rm A},{\rm B}}} \min\{\frac{\lambda_{\rm J}}{\beta},1\} +\overline{C^{B_{\rm J}=0}_{{\rm A},{\rm B}}} (1-\min\{\frac{\lambda_{\rm J}}{\beta},1\}).
\end{split}
\end{equation}
\begin{remark}
\label{remark3}
The maximum achievable average secrecy rate is $\mu_{\rm A}=  \overline{C^{B_{\rm J}>0}_{{\rm A},{\rm B}}}$ bits/sec/Hz.
If $\lambda_{\rm J}\ge \beta$, $\min\{\frac{\lambda_{\rm J}}{\beta},1\}=1$. Hence, $\mu_{\rm A}$ is constant with $\beta \le\lambda_{\rm J}\le 1$, i.e., does not change with $\lambda_{\rm J}$, and the maximum average secrecy rate is achieved. If $\lambda_{\rm J}<\beta$, $\min\{\frac{\lambda_{\rm J}}{\beta},1\}=\frac{\lambda_{\rm J}}{\beta}$ and $\mu_{\rm A}$ is linearly increasing with $\lambda_{\rm J}<\beta$.
\end{remark}

\subsubsection{The case of $\lambda_{\rm J}=1$} When Jimmy has a reliable energy supply, $\Pr\{B_{\rm J}=0\}=0$ and $\Pr\{B_{\rm J}>0\}=1$.
In this case, the average secrecy rate of the system is given by
\begin{equation} \small
\begin{split} \small
\mu_{\rm A}=  \min\{\frac{\lambda_{\rm A}}{\beta},1\} \overline{C^{B_{\rm J}>0}_{{\rm A},{\rm B}}}.
\end{split}
\end{equation}
\begin{remark}
\label{remark4}
If $\lambda_{\rm A}\ge \beta$, $\min\{\frac{\lambda_{\rm A}}{\beta},1\}=1$. Hence, $\mu_{\rm A}$ is constant with $\beta \le\lambda_{\rm A}\le 1$, and the maximum average secrecy rate is achieved, i.e., $\mu_{\rm A}=  \overline{C^{B_{\rm J}>0}_{{\rm A},{\rm B}}}$ bits/sec/Hz. If $\lambda_{\rm A}<\beta$, $\min\{\frac{\lambda_{\rm A}}{\beta},1\}=\frac{\lambda_{\rm A}}{\beta}$ and $\mu_{\rm A}$ is linearly increasing with $\lambda_{\rm A}<\beta$.
\end{remark}
\vspace{-0.5cm}
\subsection{Geo/D/1 Queueing Model}
\vspace{-0.0cm}
From \cite{krikidis,ourletter}, the probability of the Geo/D/1 energy queue with unity service rate being empty is equal to $1-\lambda_k$ for $B_k$. Applying this model to our scenario, we can rewrite (\ref{mua}) as
$
\mu_{\rm A}= \lambda_{\rm A} \left(\overline{C^{B_{\rm J}>0}_{{\rm A},{\rm B}}} \lambda_{\rm J} +\overline{C^{B_{\rm J}=0}_{{\rm A},{\rm B}}} (1-\lambda_{\rm J})\right)$. Since $\overline{C^{B_{\rm J}>0}_{{\rm A},{\rm B}}}\ge \overline{C^{B_{\rm J}=0}_{{\rm A},{\rm B}}}$, as the energy arriving at Jimmy increases, the secrecy rate increases. When Jimmy has a reliable energy supply, this represents the best-case for securing the network. In addition, the rate is linearly increasing with the average energy packet arrival rate at Alice because as $\lambda_{\rm A}$ increases, Alice will be more likely active and able to transmit data which improves her rate. The maximum average rate is achieved when $\lambda_{\rm J}=\lambda_{\rm A}=1$ energy packets/slot.
\vspace{-0.2cm}
\section{Simulation Results}
\begin{figure}
\vspace{-0.8cm}
  \centering
  \includegraphics[width=1\columnwidth]{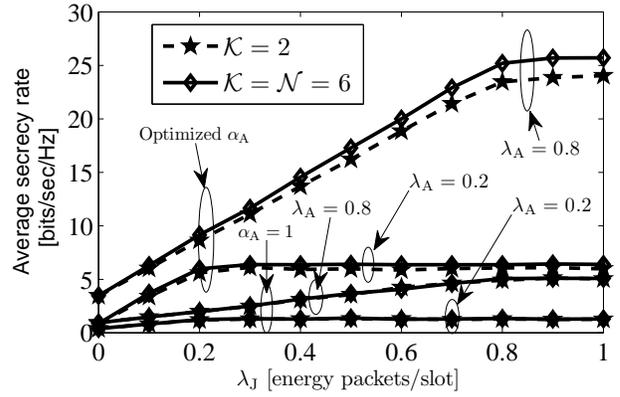}\\
  \caption{Average secrecy rate versus the energy arrival rate at Jimmy for different values of $\mathcal{K}$ and $\lambda_{\rm A}$ [energy packets/time slot].}
  \label{fig1}
\vspace{-0.4cm}
  \end{figure}

\vspace{-0.1cm}
 We simulated the system using $40000$ channel realizations and assumed that each channel coefficient is modeled as a circularly-symmetric Gaussian random variable with zero mean and unit variance. Moreover, we assume $\mathcal{N}=6$, $e_{\rm A}/\kappa/(TW)=e_{\rm J}/\kappa/(TW)=20$ dB, and $\mathcal{B}^{\max}_{\rm A}=\mathcal{B}^{\max}_{\rm J}=10$. Figure \ref{fig1} shows the average secrecy rate for our proposed jamming scheme with and without optimization over $\alpha_{\rm A}$. When we select any $\mathcal{K}$ out of the $\mathcal{N}$ antennas at Jimmy, the average secrecy rate of our proposed BF-jamming scheme is close to the case of using all of Jimmy's antennas, i.e., $\mathcal{K}=\mathcal{N}=6$, in jamming. Matching our analysis and Remarks \ref{remark3} and \ref{remark4}, the average secrecy rate increases linearly with both $\lambda_{\rm A}$ and $\lambda_{\rm J}$. If the arrival rate of a battery is high enough to saturate the battery with energy packets, the average secrecy rate becomes fixed with that arrival rate. For this reason, the curves versus $\lambda_{\rm A}$ and $\lambda_{\rm J}$ become flat with high arrival rates. The gain of $\alpha_{\rm A}$ optimization is obvious. For example, when $\lambda_{\rm A}=0.8$ energy packets/slot and $\lambda_{\rm J}=0.9$ energy packets/slot, the gain over the case of no optimization for $\alpha_{\rm A}$, i.e., $\alpha_{\rm A}=1$, is $420\%$.
\vspace{-0.15cm}
\section{Conclusions}
\vspace{-0.1cm}
In this letter, we investigated the impact of the batteries at a source node and a jammer on the achievable average secrecy rates. We showed that the average secrecy rate is nondecreasing with the arrival rates at the energy batteries and it becomes constant when these batteries are saturated with energy packets. We proposed a cooperative jamming scheme and showed that the jammer does not need to use all of its antennas for jamming Eve. The achievable performance measured by the average secrecy rate is comparable with the case of using all antennas, which requires complex hardware design since it increases the number of transmit RF chains and antennas and also complicates system design since the number of estimated channels increases. In addition, we showed that the optimization over the transmission time, $T_{\rm A}$, can significantly enhance the average secrecy rate.
\vspace{-0.1cm}
\appendices
\vspace{-0.1cm}
\section{Battery Markov Chain}
\vspace{-0.1cm}
\label{markov}
Analyzing the state balance equations of the Markov chain of the birth-death process of a Geo/Geo/1 queueing system, it is straightforward to show that the probability that the energy queue $B_k$ has $1\le \vartheta \le\mathcal{B}_k^{\max}$ energy packets, denoted by $\nu_\vartheta$, is given by
\begin{equation}
\nu_\vartheta=\nu_{\circ}\frac{1}{(1-\mu_{B_k})}\Bigg(\frac{\lambda_k (1-\mu_{B_k})}{(1-\lambda_k )\mu_{B_k}}\Bigg)^{\vartheta}=\nu_{\circ}\frac{\eta^{\vartheta}}{(1-\mu_{B_k})}
\end{equation}
where $ \vartheta\in\{1,2,\dots,\mathcal{B}_k^{\max}\}$ and $\eta=\frac{\lambda_k (1-\mu_{B_k})}{(1-\lambda_k )\mu_{B_k}}$.
Using the normalization condition $\sum_{\vartheta=0}^{\infty}\nu_\vartheta=1$, after some manipulations, the probability of $B_k$ being empty, $\nu_{\circ}$, is given by
\begin{equation}
\label{poiiiiiiiiuuuyt}
\nu_{\circ}\!=\!\frac{1}{1+\frac{1}{(1-\mu_{B_k})} \left( \frac{1-\eta^{\mathcal{B}_k^{\max}+1}}{1-\eta}-1\right)}.
\end{equation}

When $\mathcal{B}^{\max}_k$ is very large, after some mathematical manipulations, $\nu_{\circ}$ in (\ref{poiiiiiiiiuuuyt}) becomes
 \begin{equation}
\nu_{\circ}=1-\min\{\lambda_k/{\mu_{B_k}},1\}.
\end{equation}
\vspace{-0.6cm}
\bibliographystyle{IEEEtran}
 \vspace{-0.26cm}
\bibliography{IEEEabrv,bibfile}
\end{document}